\documentclass[a4paper,11pt]{amsart}
\begin{document}
\hyphenation{gra-vi-ta-tio-nal re-la-ti-vi-ty Gaus-sian
re-fe-ren-ce re-la-ti-ve gra-vi-ta-tion Schwarz-schild
ac-cor-dingly gra-vi-ta-tio-nal-ly re-la-ti-vi-stic pro-du-cing
de-ri-va-ti-ve ge-ne-ral ex-pli-citly des-cri-bed ma-the-ma-ti-cal
de-si-gnan-do-si coe-ren-za}
\title[Geometrical optics in general relativity]
{{\bf  Geometrical optics in general relativity}}
\author[Angelo Loinger]{Angelo Loinger}
\address{Dipartimento di Fisica, Universit\`a di Milano, Via
Celoria, 16 - 20133 Milano (Italy)}
\email{angelo.loinger@mi.infn.it}
\thanks{To be published on \emph{Spacetime \& Substance.}}

\begin{abstract}
General relativity includes geometrical optics. This basic fact
has relevant consequences that concern the physical meaning of the
discontinuity surfaces propagated in the gravitational field -- as
it was first emphasized by Levi-Civita.
\end{abstract}

\maketitle


\vskip0.50cm \noindent \small
\emph{\textbf{Summary}}. -- Introduction.
-- \textbf{1}. The eikonal equation. -- \textbf{2}. The
geometrical optics as a consequence of the pseudo-Riemannian
expression of the spacetime interval. -- \textbf{3}. The
geometrical optics as a consequence of Maxwell and Einstein field
equations. -- \textbf{4}. The geometrical optics according to
special relativity. -- \textbf{5}. Light propagation in ponderable
media. -- Appendix: some passages from a memoir by Levi-Civita. --

\normalsize
\vskip1.20cm \noindent \textbf{Introduction}. -- This theme has
been touched by me in several contexts, see in particular
\cite{1}, and references therein. However, on account of the
important physical consequences, it is useful to give a more
detailed treatment of it, by making explicit the involved
hypotheses.

\vskip0.50cm \noindent \textbf{1}. -- \emph{Geometrical optics}:
as it is known, the eikonal equation gives not only an
approximation -- for very small wave lengths -- of d'Alembert
equation (as Arnold Sommerfeld and Iris Runge demonstrated in 1911
\cite{2}), but yields also the \emph{exact} propagation law of the
electromagnetic wave fronts, quite independently of the values of
the wave lengths, much more, quite independently of the wave
structure -- a fundamental result, which follows from the theory
of the characteristic manifolds of the partial differential
equations of second order \cite{3}. And it is well known that the
law of propagation \emph{in vacuo} of an electromagnetic
\emph{signal} -- i.e., the law of propagation of an
electromagnetic \emph{wave front} -- is one of the basic tenet of
\emph{special relativity} (SR), whose interval $\textrm{d}s^{2}$
includes consequently the geometrical optics of light propagation
in empty space (see e.g. \cite{4}).

\vskip0.50cm \noindent \textbf{2}. -- \emph{General relativity}
(GR): consider the spacetime interval of \emph{any}
pseudo-Riemannian manifold:

\begin{equation} \label{eq:one}
\textrm{d}s^{2} = g_{jk} \, (x^{0},x^{1},x^{2},x^{3}) \,
\textrm{d}x^{j} \textrm{d}x^{k}, \quad (j,k=0,1,2,3) \quad;
\end{equation}

the metric tensor $g_{jk}$ can have \emph{any} functional
dependence on the general co-ordinates $x^{0},x^{1},x^{2},x^{3}$,
provided that (of course) the reference frame is ``appropriate''
(\emph{eigentlich}, in Hilbert's terminology \cite{5}), i.e. such
that: $g_{00}>$\nolinebreak$0$ and $g_{\alpha\beta} \, \textrm{d}x^{\alpha} \,
\textrm{d}x^{\beta}$, $(\alpha,\beta=1,2,3)$, is
negative-definite.

\par Geometrical optics according to eq.(\ref{eq:one}): we have
(cf., e.g., papers \cite{6}):

\begin{equation} \label{eq:two}
L:=g_{jk}(x)  \, \frac{\textrm{d}x^{j}}{\textrm{d}\sigma}
\frac{\textrm{d}x^{k}}{\textrm{d}\sigma}=0 \quad \textrm{--
(Lagrange-Monge)} \quad,
\end{equation}

\begin{equation} \label{eq:three}
H:=\frac{1}{2} \, g^{jk}(x)  \,
\frac{\textrm{d}z(x)}{\textrm{d}x^{j}} \,
\frac{\textrm{d}z(x)}{\textrm{d}x^{k}}=0 \quad \textrm{--
(Hamilton-Jacobi)} \quad,
\end{equation}

where $\sigma$ is a suitable parameter and $x\equiv
(x^{0},x^{1},x^{2},x^{3})$; remark that $\{$eq.(\ref{eq:two})$\}$
$\Longleftrightarrow$ $\{\textrm{d}s^{2}=0\}$. We set:

\begin{equation} \label{eq:four}
\frac{\textrm{d}x^{j}}{\textrm{d}\sigma}\equiv \dot{x}^{j} \qquad;
\qquad \frac{\partial z(x)}{\partial x^{j}}\equiv p_{j} \quad.
\end{equation}

\par Lagrange equations

\begin{equation} \label{eq:five}
\frac{\partial L}{\partial x^{j}} - \frac{\partial}{\partial
\sigma} \left( \frac{\partial L}{\partial \dot{x}^{j}} \right)=0
\end{equation}

give the characteristic lines of eq.(\ref{eq:two}), which are null
geodesics: e.m. rays. \emph{N}.\emph{B}.: eqs.(\ref{eq:five}) are
equivalent to the customary geodesic equations.

\par Hamilton equations

\begin{equation} \label{eq:six}
\dot{x}^{j} = \frac{\partial H}{\partial p_{j}} \qquad; \qquad
\dot{p}_{j} = - \frac{\partial H}{\partial x^{j}}
\end{equation}

give the characteristic lines of eq.(\ref{eq:three}) and coincide with the
null geodesics: e.m. rays. From eq.(\ref{eq:three}) we have that
$z(x)=0$ is an e.m. \emph{wave front}.

\par From the spacetime standpoint a wave front originating from a
given world point is, according to the expressive Hilbert's
terminology \cite{5}, a ``time sheath'' (\emph{Zeitscheide}).

\par \emph{These results are quite} \textbf{\emph{independent}} \emph{of the Einsteinian field
equations}. They depend only on the fact that $\textrm{d}s^{2}$ is
relative to an ``appropriate'' system of co-ordinates. They hold
even in the \textbf{\emph{absence}} of e.m. fields: an essential
property for the \emph{measurements} with ``external'' e.m.
fields.
\par Further, they are valid, in particular, if: \emph{i})
$g_{jk}$ is time independent, \emph{ii}) $g_{jk}$ is characterized
by a propagated discontinuity surface ($[$6b)$]$, see
sect.\textbf{3}.); this means that this surface is \emph{not} the
wave front of a gravitational wave, but the wave front of an
\emph{electromagnetic} radiation.

\vskip0.50cm \noindent \textbf{3}. -- Eq.(\ref{eq:three}) is also
the differential equation of the functions $z(x)$ which give the
characteristic hypersurfaces (wave fronts) $z(x)=0$ of Maxwell
field, as it was first demonstrated by Whittaker \cite{7}. And
these same functions $z(x)$ give also the characteristic
hypersurfaces $z(x)=0$ of Einstein field, as it was first
demonstrated by Levi-Civita $[$6b)$]$. This Author pointed out
explicitly that a discontinuity surface propagated in the
gravitational potential $g_{jk}$ is inevitably associated with the
conveyance of some \emph{electromagnetic} perturbation -- and
\emph{not} of a gravitational radiation.  This conclusion is
corroborated in particular by the fact that the gravitational
waves are purely mathematical undulations endowed with a
\textbf{\emph{pseudo}} energy only, and by the fact that no
motions of masses can generate gravitational waves, see \cite{8}
and references therein.

\par We see that \emph{there is a perfect concordance with the results of
sect.}\textbf{2}., which are based only on the ``appropriate''
structure of $\textrm{d}s^{2}$. In the last analysis, the reason
of this concordance is very simple: both Maxwell and Einstein
equations too are written (of course) with reference to an
``appropriate'' system of co-ordinates ($g_{00}$ positive and
$g_{\alpha\beta} \, \textrm{d}x^{\alpha} \, \textrm{d}x^{\beta}$
negative-definite).

\vskip0.50cm \noindent \textbf{4}. -- If eq.(\ref{eq:one}) is
re-interpreted as an expression of spacetime interval of
\emph{Minkowskian} world referred to a system of \emph{general}
co-ordinates $(x^{0},x^{1},x^{2},x^{3})$, sect.\textbf{2}. gives a
formulation of the geometrical optics \emph{in vacuo} as it is
described by \emph{special} relativity.

\vskip0.50cm \noindent \textbf{5}. -- A last remark. In the previous
considerations Maxwell theory is regarded from the \emph{microscopic}
 standpoint. As it is clear, this is \emph{not} a conceptual
 restriction. However, the \emph{macroscopic} e.m. theory, with the
 constitutive equations of the material, specified by various
 quantities -- \emph{in primis}, permittivity $\epsilon$ and
 magnetic permeability $\mu$ -- has a great interest of its own.
 It was thoroughly developed, according to GR, by Gordon in 1023
 \cite{9}. I limit myself to mention the following important
 result: for homogeneous (a constant $\epsilon$ and a constant
 $\mu$), insulating and uncharged ponderable media the
 differential equations of the e.m. field $F_{jk}$, $(j,k=0,1,2,3)$,
 are identical to the corresponding equations for the vacuum,
 but with the following gravitational potential $\gamma_{jk}$:

\begin{equation} \label{eq:seven}
\gamma_{jk}= g_{jk} + \left( 1- \frac{1}{\epsilon\mu}\right)
u_{j}u_{k} \quad,
\end{equation}

where $u_{j}$ is the four-velocity of matter.

\newpage
\begin{center}
\noindent \small \emph{\textbf{APPENDIX}}\normalsize
\end{center}

\vskip0.20cm \noindent
\par I reproduce here the
Introduction/Summary and the last Section of the memoir by
Levi-Civita, that concerns the characteristic hypersurfaces of the
Einsteinian field equations $[$6b)$]$.

\vskip1.00cm \noindent \par Le equazioni gravitazionali di
\textsc{Einstein} costituiscono notoriamente un sistema di dieci equazioni
alle derivate parziali del $2^{\circ}$ ordine con altrettante
funzioni incognite di quattro variabili indipendenti $x^{0},
x^{1}, x^{2}, x^{3}$. Le funzioni sono i dieci coefficienti
$g_{ik} = g_{ki}$ del

\begin{equation} \nonumber
\textrm{d}s^{2} = \sum_{0}^{3}\,_{ik} \: g_{jk} \, \textrm{d}x^{j}
\, \textrm{d}x^{k}
\end{equation}

quadridimensionale che definisce la metrica del cronotopo.

\par D'altra parte, per un sistema qualsivoglia di equazioni alle
derivate parziali, \textsc{Volterra} e \textsc{Hadamard} hanno
sviluppata una teoria generale delle caratteristiche, collegata
intimamente sia al problema d'integrazione di \textsc{Cauchy}, sia
al comportamento di eventuali superficie di discontinuit\`a, o
comunque singolari, il che, sotto l'aspetto fisico, con
riferimento al fenomeno rappresentato dalle equazioni stesse, si
interpreta come propagazione delle cosiddette fronti d'onda.

\par Naturalmente la teoria generale delle caratteristiche pu\`o in particolare
applicarsi alle equazioni gravitazionali della relativit\`a. Io mi
sono appunto proposta tale applicazione, e ne riferisco nella
presente Nota, prevalentemente preparatoria, e in altra che tosto
la seguir\`a. Si vedr\`a che, nel campo reale, le variet\`a
caratteristiche $z=0$ (ipersuperficie a tre dimensioni del
cronotopo) sono definite (con evidente significato delle notazioni)
dalla equazione alle derivate parziali del $1^{\circ}$ ordine

\begin{equation} \nonumber
H = \frac{1}{2} \, \sum_{0}^{3}\,_{ik} \: g^{jk} \, p_{i} \, p_{k} =0
\quad ,
\end{equation}

pi\`u precisamente anzi dal sussistere di tale equazione per
$z=0$, designandosi, come si vede, con $H$ la funzione
caratteristica hamiltoniana che corrisponde al $\textrm{d}s^{2}$
einsteiniano.

\par Fondamentale importanza hanno, come rilev\`o l'\textsc{Hadamard}\footnote
{Cfr. in particolare \emph{Le\c cons sur la propagation des
ondes}, Paris, Hermann, 1903, Chap. VII.}, le bicaratteristiche,
cio\`e le linee caratteristiche (secondo \textsc{Cauchy}) della equazione
del $1^{\circ}$ ordine $H = 0$, le quali (introducendo un
parametro ausiliario $t$) sono a loro volta definite dal sistema
canonico

\begin{equation} \nonumber
\frac{\textrm{d}p_{i}}{\textrm{d}t} = - \frac{\partial H}{\partial
x^{i}} \quad , \qquad \frac{\textrm{d}x^{i}}{\textrm{d}t} =
\frac{\partial H}{\partial p_{i}}
 \qquad (i= 0,1,2,3) \quad,
\end{equation}

colla specificazione che sia zero il valore di $H$ (costante per
qualsiasi soluzione del sistema). Tali linee coincidono
classicamente colle geodetiche del $\textrm{d}s^{2}$ einsteiniano;
anzi, in virt\`u di $H =0$, colle geodetiche di lunghezza nulla.

\par La conclusione appare significante, anche come riprova della perfetta
coerenza di due postulati fondamentali della relativit\`a
generale:
\par \emph{a)} il principio che la luce si propaga
secondo geodetiche di lunghezza nulla;
 \par \emph{b)} le equazioni gravitazionali che
definiscono il $\textrm{d}s^{2}$.

\par Ecco perch\`e. Con \emph{a)} si assegna la legge del moto di una generica
perturbazione luminosa. D'altra parte (\textsc{Hadamard}) le
eventuali singolarit\`a del campo, cio\`e delle equazioni
gravitazionali, si propagano secondo le bicaratteristiche delle
equazioni stesse. Infine (essendo inclusa nel $\textrm{d}s^{2}$
anche l'ottica geometrica) una superficie di discontinuit\`a che
si propaga nel campo implica altres\`i il trasporto di una qualche
perturbazione luminosa.

\par Si hanno cos\`i per la propagazione della luce due diverse
impostazioni, una fornita direttamente dal postulato \emph{a)},
cio\`e dalle geodetiche di lunghezza nulla, e l'altra derivante
dalle equazioni gravitazionali e relative bicaratteristiche. Le
due impostazioni debbono di necessit\`a condurre alle stesse
conseguenze, le quante volte coesistano effettivamente \emph{a)} e
\emph{b)} come leggi di natura. Da questo punto di vista, cio\`e
considerando la costruzione einsteiniana come definitiva, anche
nella sua struttura formale, la nostra constatazione di
coincidenza era ben prevedibile. Viceversa sotto l'aspetto
matematico, occorreva proprio dimostrare l'identit\`a, almeno nel
campo reale, delle bicaratteristiche con le geodetiche di
lunghezza nulla per rendere legittima la simultanea ammissione di
\emph{a)} e di \emph{b)}.
\\
\noindent
......................................................................................................................
\\*......................................................................................................................
\\*......................................................................................................................
\\
Terminer\`o rilevando che l'idea di applicare la teoria delle
caratteristiche alle equazioni della relativit\`a generale non \`e
nuova, essendo stata nettamente formulata dal \textsc{Whittaker}
nella sua bella \emph{Note on the law that light-rays are the null
geodesics of a gravitational field}\footnote {Proc. of the
Cambridge Phil. Society, voI. XXIV, Pt. l, 1927, pp. 32-34.}. Ivi
\`e appunto dimostrata la coincidenza delle linee di lunghezza
nulla con le bicaratteristiche delle equazioni elettromagnetiche
(di \textsc{Maxwell}, adattate al $\textrm{d}s^{2}$ cronotopico),
ed \`e pur messo in luce che non poteva essere altrimenti (in una
teoria conseguente), in quanto le equazioni elettromagnetiche
includono anche l'ottica ondulatoria, e, come caso limite,
l'ottica geometrica.

\par La precedente trattazione \`e in certo modo complementare a
quella del \textsc{Whittaker}, poich\'e concerne le equazioni
gravitazionali ed il loro collegamento diretto coll'ottica
geometrica (indipendentemente da ogni teoria dei fenomeni
elettromagnetici).

\end{document}